# Gryphon: An Information Flow Based Approach to Message Brokering

Robert Strom, Guruduth Banavar, Tushar Chandra, Marc Kaplan, Kevan Miller, Bodhi Mukherjee,
Daniel Sturman, and Michael Ward
*IBM TJ Watson Research Center, 30 Saw Mill River Road, Hawthorne, NY 10532*
*Contact E-mail: banavar@watson.ibm.com*

Gryphon is a distributed computing paradigm for *message brokering*, which is the transferring of information in the form of streams of events from information providers to information consumers. This abstract outlines the major problems in message brokering and Gryphon's approach to solving them.

In Gryphon, the flow of streams of events is described via an *information flow graph*. The information flow graph specifies the selective delivery of events, the transformation of events, and the generation of derived events as a function of *states* computed from event histories. For this, Gryphon derives from and integrates the best features of distributed communications technology and database technology.

Message brokering is motivated by the need for efficient delivery of information across a large number of users and applications, in an environment characterized by heterogeneity of computing platforms, anonymity between information producers and consumers, and dynamic change due to system evolution. Within a single business, such as a stock exchange or a weather forecasting agency, there is a dynamically varying number of sub-applications supplying events, and a varying number consuming events. The suppliers and consumers may not necessarily be aware of one another; instead the suppliers may simply be supplying information of a certain type to any interested consumer and each consumer may be interested in subsets of this information having particular properties. For example, in a stock exchange, one consumer may be interested in all stock trades greater than 1000 shares, and another in specific market trends, such as all stock trades representing a drop of more than 10 points from the previous day's high.

There is also a growing need to "glue" together applications within multiple businesses, to support inter-business network commerce or maybe as a result of mergers and acquisitions. For example, a retailer may need to connect to its suppliers and customers, or a customer to various retailers and financial organizations. This may require transforming events from different sources into a compatible form, merging them, and selecting from these events.

Message brokering is an extension of *publish-subscribe* technology [Powell96]. The Gryphon approach augments the publish-subscribe paradigm with the following features:

- *Content-based subscription,* in which events are selected by predicates on their content rather than by pre-assigned subject categories;
- *Event transformations*, which convert events by projecting and applying functions to data in events;
- *Event stream interpretation*, which allows sequences of events to be collapsed to a state and/or expanded back to a new sequence of events; and
- *Reflection*, which allows system management through *meta-events*.

Gryphon technology includes a collection of efficient implementations to support this paradigm and still provide scalability, high throughput and low latency.

## 1. The Gryphon Model

As mentioned earlier, event processing in Gryphon is described via an information flow graph. An information flow graph is a directed acyclic graph constituting an abstraction of the flow of events in the system. In the example shown in the figure below, stock trades from two information sources, NYSE and NASDAQ, are combined, transformed, filtered and delivered to a client. The two sources produce events of type [price, volume], which are merged into a single stream. The arc labeled *transform* computes a new stream of events of type [capital], and the arc labeled *select* filters out events with capital less than $1,000,000.

A Gryphon information flow graph is an abstraction because Gryphon is free to physically implement the flow any way it chooses, possibly radically altering the flow pattern, provided that the consumers see the appropriate streams of events consistent with the incoming events and the transformations specified by the graph. Gryphon optimizes graphs and deploys them over a network of *brokers* (or servers). The broker network is responsible for handling client connections and for distributing events.

The nodes of the graph are called *information spaces*. Information spaces are either *event histories* --

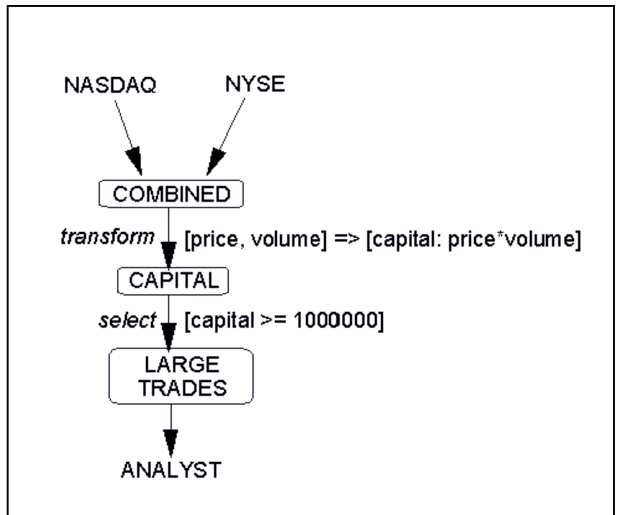

monotonically growing collections of events, e.g., stock trade events of the form [price, volume], or *event interpretations* -- states representing interpretations of sequences of events, such as a table [latestprice, highestprice]. Each information space has a *schema* defining the structure of the history or state it contains.

Each arc in the information flow graph defines an operation to derive the information space at the head from that at the tail. Arcs can be one of the following types:

- *select(P)* specifies that the destination event history contains the subset of events in the source event history that satisfy the predicate *P*. The two event histories have the same schema.
- *transform(T)* specifies that each event in the destination event history is obtained by applying function *T* to the corresponding event in the source event history.
- *merge* combines two or more event histories of the same schema into a single history. This operation is implicit when multiple arcs lead to the same information space.
- *interpret(I)* converts a source event history to a destination state by applying an interpretation function *I* to the history. Each time a new event arrives, this interpretation will be (incrementally) re-evaluated.
- *expand(I)* The inversion of *interpret*: converts a state to an event history which is equivalent to that state under function *I*. This is a non-deterministic function: in particular, interpreting an event history and re-expanding it with the same *I* may yield the identical event history, but may also yield a different history which yields an equivalent state under *I*.

## 2. System Research Directions

The Gryphon system consists of several components for efficiently realizing the information flow graph over a large distributed network of brokers. In particular, our current efforts are addressing the following technologies and their integration:

*Event matching* -- determining, for each event, the subset of N subscriptions that match the event. Our current algorithm for event matching grows sub-linearly with N [Strom+98].

*Multicasting* -- routing the events from source to all destinations while avoiding unnecessary processing of messages at brokers and long message headers on events [Strom+98]. Existing multicast techniques from the literature use the concept of groups [Birman93], and do not apply to content-based pub/sub systems.

*Graph transformations* -- reordering the selects, transforms, and interpretations to minimize the number of events sent through the network and the load on the brokers.

*Fault-tolerance* -- preserving the appearance of a persistent information flow graph in the presence of failures. In addition, guaranteeing that, when required, clients have consistent views of information spaces even in the presence of failure so that some clients don't see that an information space contains a message while others see that the same information space has lost the message.

*Ordered delivery* -- guaranteeing that, when required, clients have consistent view of the order of events in an information space.

*Optimistic delivery* -- when a client's view of an information space is through an interpretation, exploits the non-determinism of the equivalent state to deliver messages to the client early, out-of-order, or to drop messages.

*Compression* -- when a client's view of an information space is through an interpretation and the client disconnects and reconnects, exploits the non-determinism of the equivalent state to deliver a compressed sequence of events that captures the same state.

*Reconfiguration* -- allows the physical broker network to be dynamically updated and extended without disturbing the logical view of a persistent information flow graph.

*Reflection* -- capturing events corresponding to requests to change the information flow graph, and confirmed changes to the information flow graph in a special *meta-event space*.

*Security* -- dealing with issues concerning the lack of full mutual trust between domains in the physical broker network.